\documentclass[10pt,aps,prb,twocolumn,floatfix,showpacs,superscriptaddress,longbibliography]{revtex4-1}

\usepackage{graphicx,amsmath,amsfonts,bm}
\usepackage{tabularx}
\usepackage[colorlinks=true, citecolor=blue, linkcolor=blue, urlcolor=blue]{hyperref}
\usepackage[all]{hypcap}

\newcolumntype{Y}{>{\centering\arraybackslash}X}

\graphicspath{{figs/}}

\DeclareMathOperator{\diag}{diag}

\begin{document}

\title{Transmission in graphene-topological insulator heterostructures}
\author{C. De Beule}
\email{christophe.debeule@uantwerpen.be}
\affiliation{Department of Physics, University of Antwerp, 2020 Antwerp, Belgium}
\author{M. Zarenia}
\affiliation{Department of Physics, University of Antwerp, 2020 Antwerp, Belgium}
\author{B. Partoens}
\affiliation{Department of Physics, University of Antwerp, 2020 Antwerp, Belgium}
\begin{abstract}
We investigate scattering of the topological surface state of a three-dimensional time-reversal invariant topological insulator when graphene is deposited on the  topological-insulator surface. Specifically, we consider the (111) surface of a Bi$_2$Se$_3$-like topological insulator. We present a low-energy model for the bulk graphene-topological insulator heterostructure and we calculate the transmission probability at zigzag and armchair edges of the deposited graphene, and the conductance through graphene nanoribbon barriers and show that its features can be understood from antiresonances in the transmission probability.
\end{abstract}

\maketitle

\section{Introduction}

Topological insulators \cite{2dteo1,2dteo2,2dexp1,2dexp2,3dteo1,3dteo2,3dexp} (TIs) are materials with metallic surface states that are topologically protected by time-reversal symmetry and the insulating bulk. In the simplest case, the topological surface state is given by a single Dirac cone that is characterized by spin-momentum locking \cite{rev1,rev2}. The topological surface states have potential applications in spintronics and quantum computation, and it is therefore desirable to tune their properties to suit specific needs. Tailoring the surface states can also lead to new physics. For example, by changing their dispersion relation \cite{zarenia2014, jang2015} they become more susceptible towards interactions which could lead to novel strongly correlated phases.

One possibility consists of depositing a thin layer of a non-topological metal on the topological-insulator surface (TIS), effectively changing the boundary conditions at the surface \cite{wang2013, shoman2015, bian2016}. The topological surface state migrates to the new surface obtaining different properties depending on the type of deposited thin film. In particular, graphene, is a very interesting candidate, for a number of reasons. Graphene has been studied extensively in the last decade and its properties are well known: It hosts four Dirac cones in its bulk whose Dirac structure act on the sublattice pseudospin of the honeycomb lattice \cite{reviewGraphene}. The interplay between the Dirac cones of graphene and the topological Dirac cone can drastically change the properties of the resulting topological surface state \cite{zhang14}. Moreover, the lattice mismatch between graphene and the natural surface of several TIs is very small, from a few percent to near perfect matching.

In this work, we  investigate transmission in heterostructures made from depositing graphene on top of the (111) surface of a Bi$_2$Se$_3$-like TI. This setup was recently experimentally realized \cite{3dexpTI}. The archetypal strong topological insulator, Bi$_2$Se$_3$, has a layered crystal structure where each layer has trigonal symmetry and the layers are generally only weakly coupled by van der Waals-like bonding. The (111) surface is parallel to these layers and hosts a single Dirac cone at the center of the surface Brillouin zone (BZ). If graphene is placed on top of this surface in the commensurate $\sqrt{3}{\times}\sqrt{3}$ R30 stacking configuration, the graphene Dirac cones are folded onto the topological Dirac cone so that even weak coupling can strongly affect the low-energy physics if the chemical potential is tuned accordingly \cite{zhang14}. In this configuration, the trigonal lattice of graphene and the TIS are rotated by $30^\circ$ with respect to each other and the surface unit cell contains six carbon atoms from graphene and one atom from the TIS.
The most promising currently known TIs for realizing such a heterostructure are Sb$_2$Te$_3$, which has recently been fabricated \cite{bian2016}, and TlBiSe$_2$ \cite{TlBiSe2prl1,TlBiSe2prl2,TlBiSe2prl3}. Both have only a lattice mismatch of the order of $0.1$\% \cite{lattice2,TlBiSe2prl3}. While the interlayer coupling of Sb$_2$Te$_3$ is van der Waals-like, that of TlBiSe$_2$ is more covalent \cite{TlBiSe2prl1}, allowing for stronger coupling between graphene and the TIS in the latter case. In Table \ref{tab:materials}, we show a list of potential TIs together with the lattice mismatch, the band gap, and the Fermi velocity of the topological Dirac cone.



The paper is further organized as follows: In Sec.\ \ref{section:model}, we introduce the model for the graphene-topological insulator heterostructure. We consider different stacking configurations, elucidate the physics by block diagonalizing the Hamiltonian, and derive a low-energy model. In Sec.\ \ref{section:transmission}, we solve the two-dimensional scattering problem for different geometries. In particular, we consider the interface between the bare TIS and the heterostructure for both zigzag and armchair graphene edges. We also consider barriers consisting of graphene nanoribbons deposited on top of the TIS where the bare TIS acts as leads. We discuss our results for the transmission probability, the bound states, and the conductance through the different barriers in Sec.\ \ref{section:results} and present the summary and conclusions of the paper in Sec.\ \ref{section:summary}.
\begin{table}
	\centering
	\begin{tabularx}{\linewidth}{ X Y Y Y }
	\toprule
	& mismatch (\%) & gap (eV) & $v_s/v_g$ \\
	\hline
	Bi$_2$Se$_3$ &  2.7 \cite{lattice1} & 0.3 \cite{Bi2Se3} & 0.5 \cite{Bi2Se3}, 0.3 \cite{TlBiSe2prl3} \\
	Sb$_2$Te$_3$ & 0.1 \cite{Sb2Te3} & 0.3 \cite{Sb2Te3} & 0.4 \cite{Sb2Te3} \\ 
	Bi$_2$Te$_2$Se & 0.9 \cite{lattice1} & 0.3 \cite{Bi2Te2Se} & 0.5 \cite{Bi2Te2Se} \\ 
	TlBiSe$_2$ & 0.2 \cite{lattice2} & 0.35 \cite{TlBiSe2prl1}, 0.3 \cite{TlBiSe2prl2}, 0.2 \cite{TlBiSe2prl3} & 0.3 \cite{TlBiSe2prl1}, 0.4  \cite{TlBiSe2prl2}, 0.7 \cite{TlBiSe2prl3} \\ 
	\toprule
	\end{tabularx}
	\caption{The lattice mismatch of the graphene-TI heterostructure, band gap, and Fermi velocity $v_s$ for some TIs with a simple Dirac cone. We have taken $a = 2.46$ \r{A} and $v_g = 10^6$ m/s for the lattice constant and Fermi velocity of graphene, respectively \cite{reviewGraphene}.}
	\label{tab:materials}
\end{table}

\section{Model} \label{section:model}

\begin{figure}
	\centering
	\includegraphics[width=\linewidth]{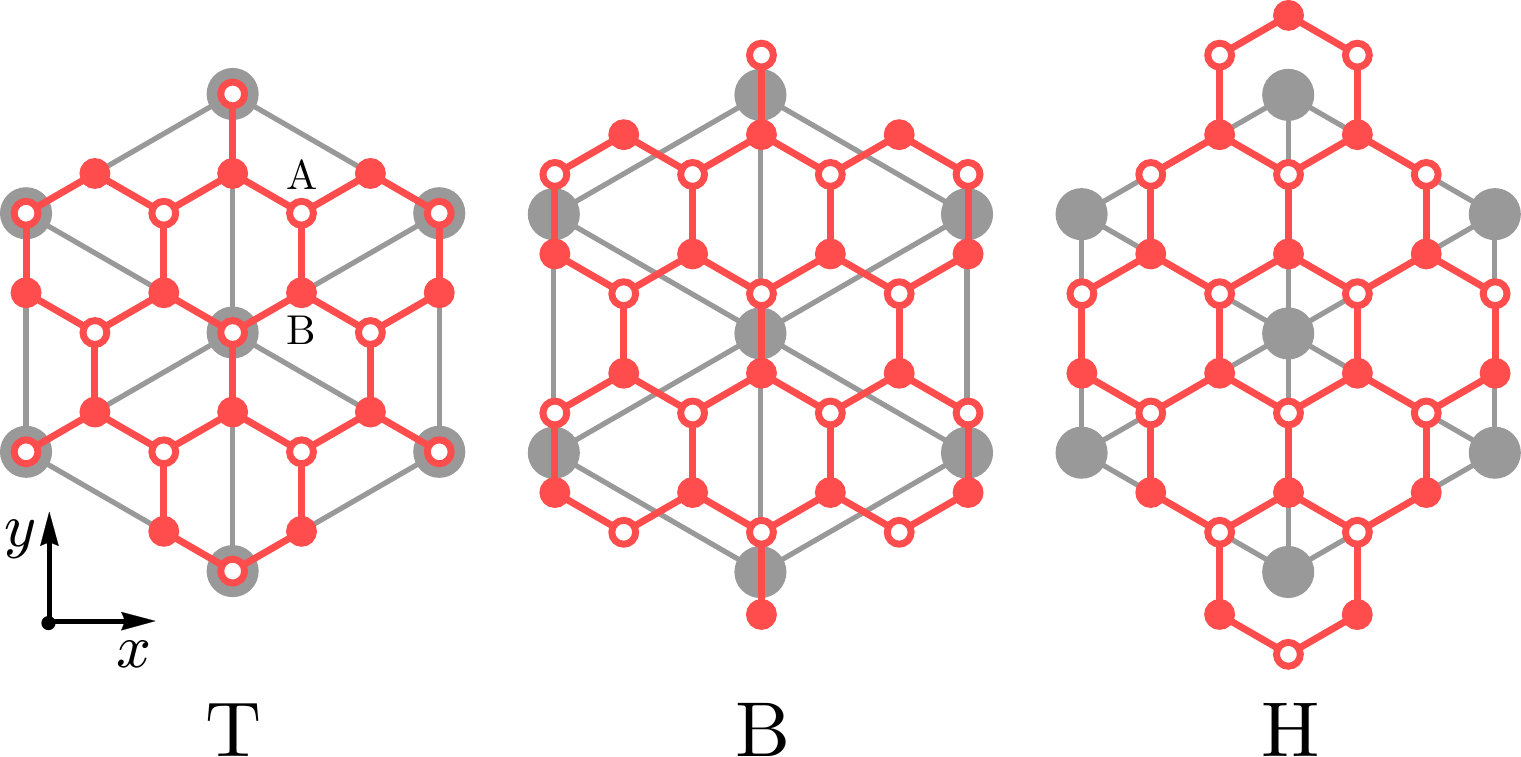}
	\caption{(Color online) Top view of the different commensurate $\sqrt{3}{\times}\sqrt{3}$ R30 stacking configurations of the graphene (red) and topological-insulator surface (gray) heterostructure. The structures differ by the position of the TIS atom in the unit cell: (T) one sublattice on top, (B) bond on top, and (H) in the center of a graphene hexagon.}
	\label{fig:stacking}
\end{figure}
We consider the surface of a Bi$_2$Se$_3$-like time-reversal invariant strong topological insulator on which a monolayer of graphene is deposited. The Hamiltonian reads
\begin{equation}
H = H_G + H_{\rm TIS} + V,
\end{equation}
where $H_G$ and $H_{\rm TIS}$ are, respectively, the Hamiltonians of graphene and the topological-insulator surface and $V$ represents the coupling between them.

For commensurate $\sqrt{3}{\times}\sqrt{3}$ R30 stacking, illustrated in Fig.\ \ref{fig:stacking}, the Dirac cones at the $K$ and $K'$ point of graphene are folded onto the zone center $\bar \Gamma$ of the TIS BZ which harbors the topological Dirac cone. Hence, the low-energy Bloch Hamiltonian becomes
\begin{equation} \label{eq:ham}
h(\bm k) =
\begin{pmatrix}
h_{K} & 0 & \mathcal V^\dag \\
0 & h_{K'} & \mathcal V^\dag \\
\mathcal V & \mathcal V & h_{\rm TIS}
\end{pmatrix},
\end{equation}
where $\mathcal V$ are the coupling matrix elements of $V$ between the $p_z$ orbitals of graphene and the TIS.
In the coordinate system shown in Fig.\ \ref{fig:stacking}, we have
\begin{align}
h_{K}(\bm k) & = \hbar v_g s_0 \otimes \left( \bm \sigma \cdot \bm k \right) - \mu \label{eq:hamK1} \\
h_{K'}(\bm k) & = \hbar v_g s_0 \otimes \left( -\bm \sigma^* \cdot \bm k \right) - \mu \label{eq:hamK2} \\
h_{\rm TIS}(\bm k) & = \hbar v_s \left( \bm{\hat z} \times \bm s \right) \cdot \bm k, \label{eq:hamT} 
\end{align}
where $v_g$ and $v_s$ are respectively the Fermi velocity of graphene and the bare TIS, $\mu$ is the chemical potential difference between graphene and the TIS, and $\bm \sigma$ and $\bm s$ are the Pauli matrices corresponding to pseudospin and spin, respectively. In the remainder of this article, we put $\hbar = 1$ unless otherwise stated.

In our basis, the time-reversal operator becomes
\begin{equation}
\Theta = \left( \tau_x\otimes i s_y \otimes \sigma_0 \right) \oplus i s_y \mathcal K,
\end{equation}
where $\mathcal K$ denotes complex conjugation and $\tau_x$ is the Pauli matrix in valley space. Time-reversal symmetry gives $\Theta h(-\bm k) \Theta^{-1} = h(\bm k)$ and constrains the coupling $\mathcal V$:
\begin{equation}
\mathcal V(\bm k) = \begin{pmatrix} t_A(\bm k) & t_B(\bm k) & \lambda_A(\bm k) & \lambda_B(\bm k) \\ -\lambda_A(-\bm k)^* & -\lambda_B(-\bm k)^* & t_A(-\bm k)^* & t_A(-\bm k)^* \end{pmatrix},
\end{equation}
where $t_A$ and $t_B$ correspond to coupling between the same spins, and $\lambda_A$ and $\lambda_B$ to coupling between different spins. We do not consider the latter and hence we put $\lambda_A = \lambda_B = 0$. The form of $t_A$ and $t_B$ depends on the specific stacking: In Fig.\ \ref{fig:stacking}, we show the three most symmetrical stacking configurations. \emph{Ab initio} studies on graphene deposited on thin films of Sb$_2$Te$_3$ show that the binding energy of these structures only differ by a few meV with H the most stable configuration \cite{jin13}.

For the T and B structure shown in Fig.\ \ref{fig:stacking}, the coupling is given, in lowest order, by
\begin{equation}
\mathcal V = \begin{pmatrix} t_A & t_B & 0 & 0 \\ 0 & 0 & t_A & t_B \end{pmatrix},
\end{equation}
where $t_A$ ($t_B$) is the coupling matrix element between the TIS and the A (B) sublattice. Specifically, in lowest order, we have $t_B = 0$ for T stacking and $t_A = t_B$ for B stacking. However, for the $H$ structure, also shown in Fig.\ \ref{fig:stacking}, the lowest-order coupling vanishes at $\bm k = 0$.

\begin{figure}
	\centering
	\includegraphics[width=\linewidth]{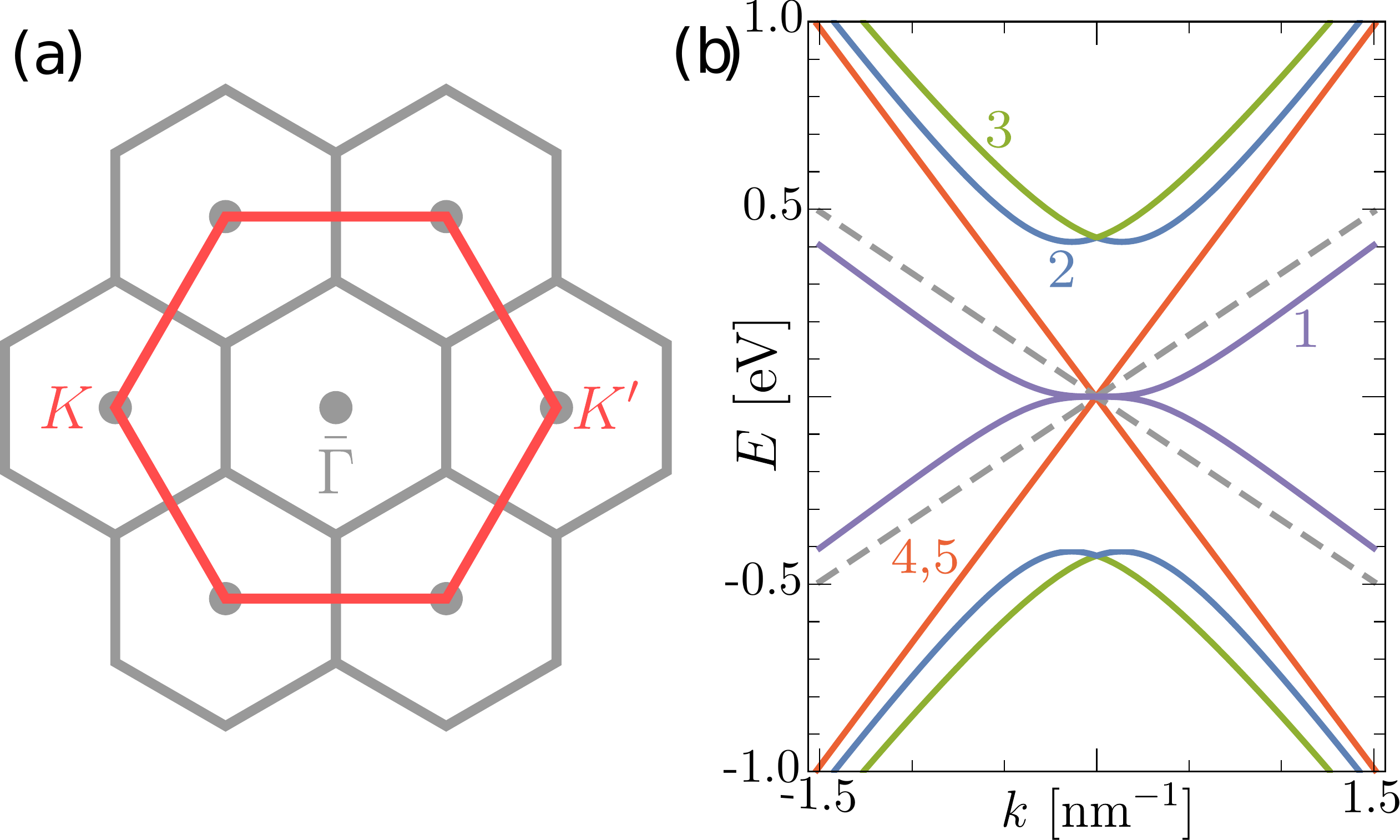}
	\caption{(Color online) (a) Momentum space of the commensurate $\sqrt{3}{\times}\sqrt{3}$ R30 stacking configuration shown in Fig.\ \ref{fig:stacking} in the extended zone scheme. The small (gray) hexagons correspond to the TIS, where the dots are reciprocal lattice points, and the large (red) hexagon is the first BZ of graphene;  the $K$ and $K'$ point of graphene are folded to the $\bar \Gamma$ point of the surface BZ. (b) Energy spectrum of the T structure with $\mu =0$, $t_A=0.3$ eV, $t_B=0$, and $v_s = v_g/2$. The dashed curve is the original topological Dirac cone and the index $n=1,\ldots,5$ refers to the $n$-th scattering channel.}
	\label{bands1}
\end{figure}

The energy spectrum of the T structure is shown in Fig.\ \ref{bands1} for $\mu = 0$. A similar energy spectrum is obtained for the B structure. For the H structure, the spectrum only shows a change in the Fermi velocity of the Dirac cones. The spectrum shown in Fig.\ \ref{bands1} is thus generic for any $\sqrt{3}{\times}\sqrt{3}$ R30 stacking configuration at low energies with the exception of H stacking. Since we are interested in strong coupling between the Dirac cones, we restrict ourselves to the T structure with $\mu=0$. Thus, we put $t_A = t$ and $t_B = 0$ in the remainder of the article.

\subsection*{Valley exchange}

From the energy spectrum for the T structure, shown in Fig.\ \ref{bands1}, we observe that two of the four Dirac cones of graphene do not couple at all with the TIS. This suggests that the graphene Dirac cones partly decouple. The symmetry that enables this block diagonalization is \emph{valley exchange}: $K \leftrightarrow K'$. States that are even under valley exchange couple to the TIS, while states that are odd under valley exchange do not. Formally, we can write
\begin{equation} \label{eq:block}
UhU^\dag = h_+ \oplus h_-, 
\end{equation}
where $U=U(\bm k)$ is a suitable unitary transformation, whose explicit form is given in the Appendix for T stacking. This is illustrated in Fig. \ref{fig:scheme}. For T stacking, the two blocks $h_+$ and $h_-$ can be written as
\begin{align}
h_+ & =
\begin{pmatrix} 
0 & -v_g k_- & & & & \\
-v_g k_+ & 0 & \sqrt{2}t & & & \\
& \sqrt{2}t & 0 & v_s ik_- & & \\
& & -v_s ik_+ & 0 & \sqrt{2}t & \\
& & & \sqrt{2}t & 0 & v_g k_- \\
& & & & v_g k_+ & 0 \\
\end{pmatrix} \label{eq:block1} \\[2mm]
h_- & = v_g \left( \bm{\sigma} \cdot \bm{k} \oplus -\bm{\sigma}^* \cdot \bm{k}  \right), \label{eq:block2}
\end{align}
with $k_\pm = k_x \pm ik_y$. We find that $h_+$ is equivalent to the low-energy Hamiltonian of spinless ABC-stacked trilayer graphene for which the middle layer is triaxially strained, while $h_-$ is like a spinless version of graphene \cite{trilayer}. We can understand the decoupling as follows: The matrix elements between the odd subspace and the topological surface state pick up a minus sign under time reversal, so that they have to be zero because the coupling is time-reversal invariant.
\begin{figure}
	\centering
	\includegraphics[width=.8\linewidth]{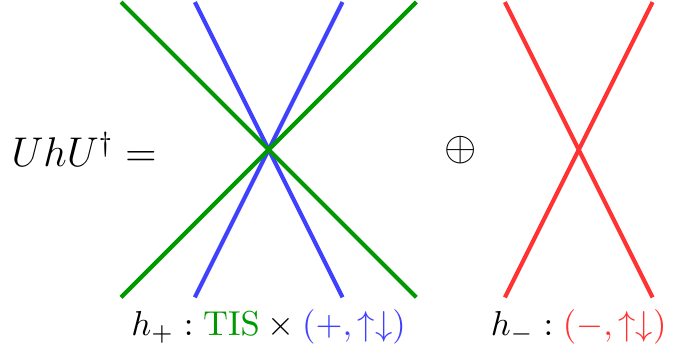}\hspace{0.5cm}
	\caption{(Color online) Graphical representation of the block diagonalization of $h(\bm k)$ into subspaces that are even ($+$, blue cone) and odd ($-$, red cone) under valley exchange. The spectra are shown for $t=0$. Only the even subspace couples to the topological-insulator surface (green cone).}
	\label{fig:scheme}
\end{figure}

In analogy with ABC trilayer graphene, the energy dispersion is cubic at low energies ($v k/t \ll 1$) \cite{trilayer}. Moreover, we find that the topological surface state migrates to the graphene. For our choice of unitary transformation, the effective low-energy Hamiltonian close to the $\bar{\Gamma}$ point becomes
\begin{equation} \label{eq:hameff}
\frac{v_g^2v_s}{2t^2}
\begin{pmatrix}
0 & k_-^3 \\ k_+^3 & 0
\end{pmatrix} \oplus h_-,
\end{equation}
where the basis of the first $2\times2$ block is $\{i\left| \psi_B^+ \uparrow \right>,\left| \psi_B^+ \downarrow \right>\}$. The $+$ indicates that these states are symmetric-like superposition of $K$ and $K'$ which are given explicitly in the Appendix. Note that these states correspond to the sublattice that does not couple directly with the TIS in lowest order of $v k/t$. Accordingly, the low-energy physics is understood in terms of an intermediate virtual process: In lowest order, the spin states of the $B^+$ sublattice couple to each other via the $A^+$ sublattice and the original topological surface state, leading to the cubic dispersion. Apart from the cubic dispersion, two uncoupled valley odd cones remain. The presence of boundaries, however, can induce coupling to these cones and they are not robust against time-reversal invariant perturbations in general. 
Similarly, an AB-stacked graphene bilayer that is suitably deposited on the TIS leads to a quintic dispersion at low energies, now localized on a single sublattice of the top layer of the bilayer, together with two quadratic cones corresponding to the odd subspace of the bilayer \cite{zhang14}.

\begin{figure}
	\centering
	\includegraphics[width=0.6\linewidth]{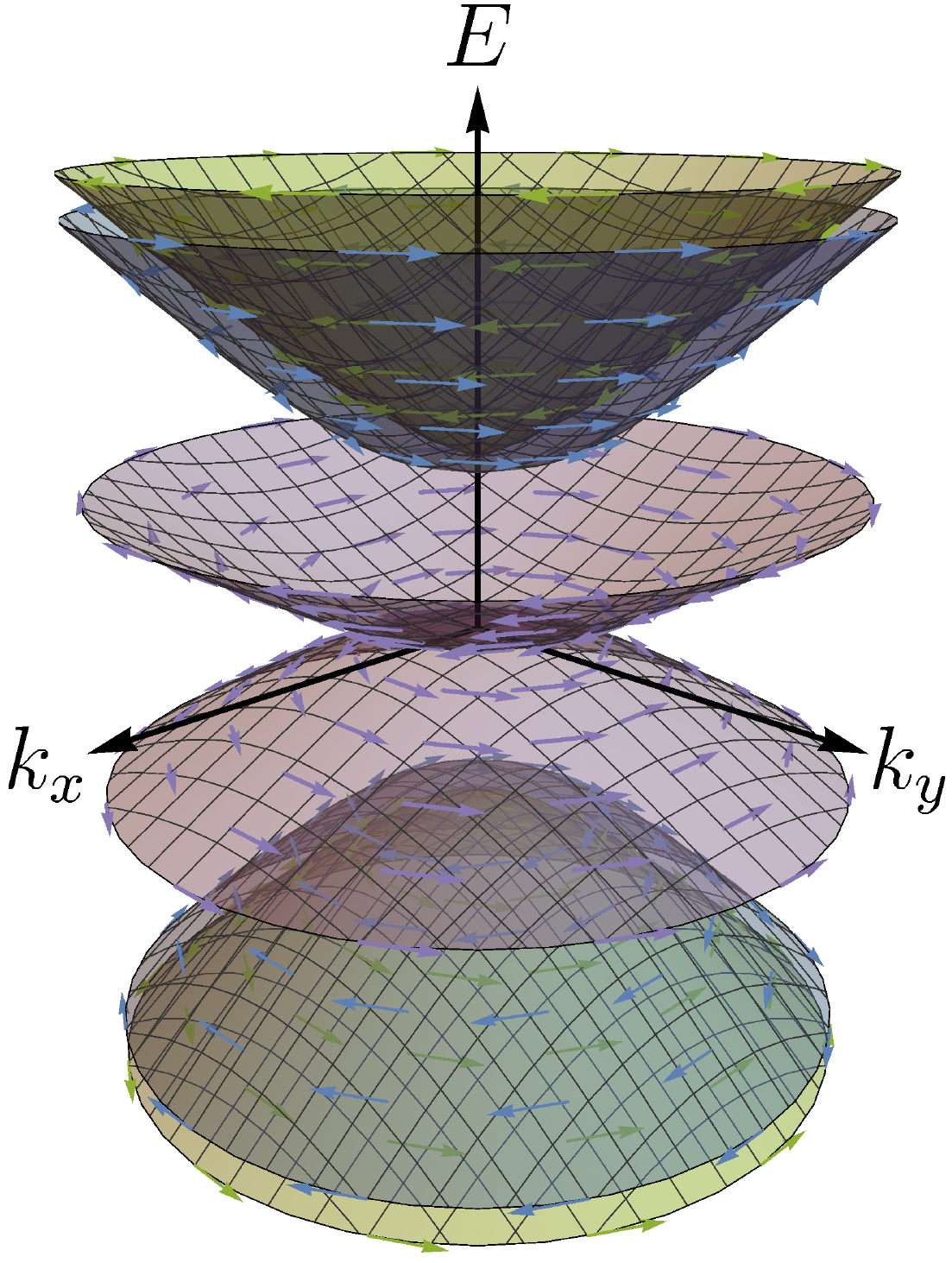}
	\caption{(Color online) Low-energy spectrum for T (or B) stacking where the corresponding spin expectation values are shown as arrows. All bands except the two valley odd Dirac cones that decouple in the bulk are shown.}
	\label{fig3}
\end{figure}

In Fig. \ref{fig3}, we show the two-dimensional bands obtained from $h_+$ together with the corresponding spin expectation value. While the decoupled Dirac cones from $h_-$ remain $s_z$ eigenstates, the other bands inherit their spin structure from the original topological surface state. Besides the cubic Dirac bands, there are two bands from the valley even subspace that have a Rashba-like dispersion with opposite spin-momentum locking. These states arise from proximity-induced Rashba coupling since reflection symmetry about the graphene plane is broken when deposited on the TIS. By expanding the dispersion relation to second order in $k$, we find that the Rashba momentum and energy splitting are approximately given by $\displaystyle{(2 \sqrt{2} t v_s)/(4v_g^2+v_s^2)}$ and $\displaystyle{(t v_s^2)/[\sqrt{2} \left( 4v_g^2 + v_s^2 \right)]}$.

\section{Transmission} \label{section:transmission}

In this section, we consider elastic scattering of the topological surface state at a graphene-topological insulator heterostructure for the T structure. First, we consider scattering at a graphene step terminated by zigzag or armchair edges, where an incident wave on the bare TIS coming in from the left ($x<0$) is transmitted to the right ($x>0$) into a semi-infinite region of the heterostructure. Next, we consider transmission through a graphene nanoribbon barrier of finite width.

We work in the original basis in which the Hamiltonian takes the form given in Eq.\ \eqref{eq:ham}. In the basis where the Hamiltonian is block diagonal, the boundary conditions at a graphene edge can couple the two blocks and we prefer to work in the original basis where the boundary conditions are straightforward.

If we take the coordinate system shown in Fig.\ \ref{fig:geometry}, the scattering state for the bare TIS is given by an incident and reflected wave
\begin{equation} \label{eq:sol1}
\Phi_I(x) = \phi_i e^{ik_x x} + r \phi_r e^{-ik_x x},
\end{equation}
where $r$ is the reflection coefficient and
\begin{equation} \label{eq:spinor1}
\phi_i = \begin{pmatrix} E/v_s \\ k_y - ik_x \end{pmatrix}, \quad \phi_r = \begin{pmatrix} E/v_s \\ k_y + ik_x \end{pmatrix},
\end{equation}
are the corresponding spinors with $E$ the energy relative to the Dirac point. We have left out normalization constants since they are irrelevant for our calculation. The longitudinal and transverse momentum are given by $k_x$ and $k_y$, respectively. The latter is conserved because of translation symmetry in the $y$ direction. The longitudinal momentum is given by
\begin{equation} \label{eq:mom1}
k_x = \textrm{sign}(E) \sqrt{(E/v_s)^2 - k_y^2},
\end{equation}
where $E = v_s k$ for the Dirac cone of the TIS. The sign of $k_x$ makes sure that the incident wave propagates to the right and the reflected wave propagates to the left.

\subsection{Graphene step}

\subsubsection{Scattering states}

In the semi-infinite T region, the wave function can be written as
\begin{equation} \label{eq:sol2}
\Phi_{II}(x) = \sum_{n=1}^5 t_n \psi_n e^{iq_{nx} x},
\end{equation}
where $t_n$, $\psi_n$, $\bm q_n = q_{nx} \bm{\hat x} + k_y \bm{\hat y}$ are, respectively, the transmission coefficient, the spinor, and the momentum of the $n$th scattering channel of the heterostructure. The sign of $q_{nx}$ is chosen such that for scattering modes the group velocity is positive and the wave propagates to the right, while for evanescent modes it is chosen such that the imaginary part is positive since otherwise the solution from Eq.\ \eqref{eq:sol2} would blow up for $x\rightarrow \infty$. The bands corresponding to the different transmission channels are shown in Fig.\ \ref{bands1}: $\psi_1$ corresponds to the cubic dispersion, $\psi_2$ and $\psi_3$ to the Rashba-like bands, while $\psi_4$ and $\psi_5$ correspond to the two uncoupled Dirac cones.
Scattering to a particular channel only takes place if $q_x$ is real, otherwise the corresponding wave function is evanescent and does not contribute to transmission. We also expect that there is no transmission into the channels $\psi_4$ and $\psi_5$ that are decoupled from the TIS in the bulk. The presence of certain boundaries, however, allows for transmission to $\psi_4$ and $\psi_5$, as we show below.

The spinors $\psi_4$ and $\psi_5$ can be explicitly written as
\begin{equation} \label{eq:oddmodes}
\psi_4 = \begin{pmatrix} E/v_g \\ q_{4x} + ik_y \\ 0 \\ 0 \\ -E/v_g \\ q_{4x} - ik_y \\ 0 \\ 0 \\0 \\ 0 \end{pmatrix}, \quad \psi_5 =  \begin{pmatrix} 0 \\ 0 \\ E/v_g \\ q_{5x} + ik_y \\ 0 \\ 0 \\ -E/v_g \\ q_{5x} - ik_y \\ 0 \\ 0 \end{pmatrix},
\end{equation}
with
\begin{equation} \label{eq:mom2}
q_{4x} = q_{5x} = \textrm{sign}(E) \sqrt{(E/v_g)^2 - k_y^2}.
\end{equation}
It is clear that the spinors $\psi_4$ and $\psi_5$ are $s_z$ eigenstates and have odd valley parity since they are antisymmetric superpositions of states at $K$ and $K'$. The other spinors $\psi_1$, $\psi_2$, and $\psi_3$ and the corresponding wave vectors are found numerically. The secular equation $|H(q_x,k_y)-E|=0$ yields a bicubic equation:
\begin{equation} \label{eq:mom3}
\begin{aligned}
& \frac{v_g^4 v_s^2}{E^2} q_m^6 - v_g^2 \left( v_g^2 + 2 v_s^2 \right) q_m^4 \\
& + \left[ \left( 2 v_g^2 + v_s^2 \right) E^2 - 4 v_g^2 t^2\right] q_m^2 - \left( E^2 - 2 t^2 \right)^2 = 0,
\end{aligned}
\end{equation}
where $q_{mx} = \pm \sqrt{q_m^2 - k_y^2}$ with $m=1,2,3$. The sign is determined so that scattering modes propagate to the right and evanescent modes decay inside the T region.

\subsubsection{Boundary conditions}

\begin{figure}
	\centering
	\includegraphics[width=\linewidth]{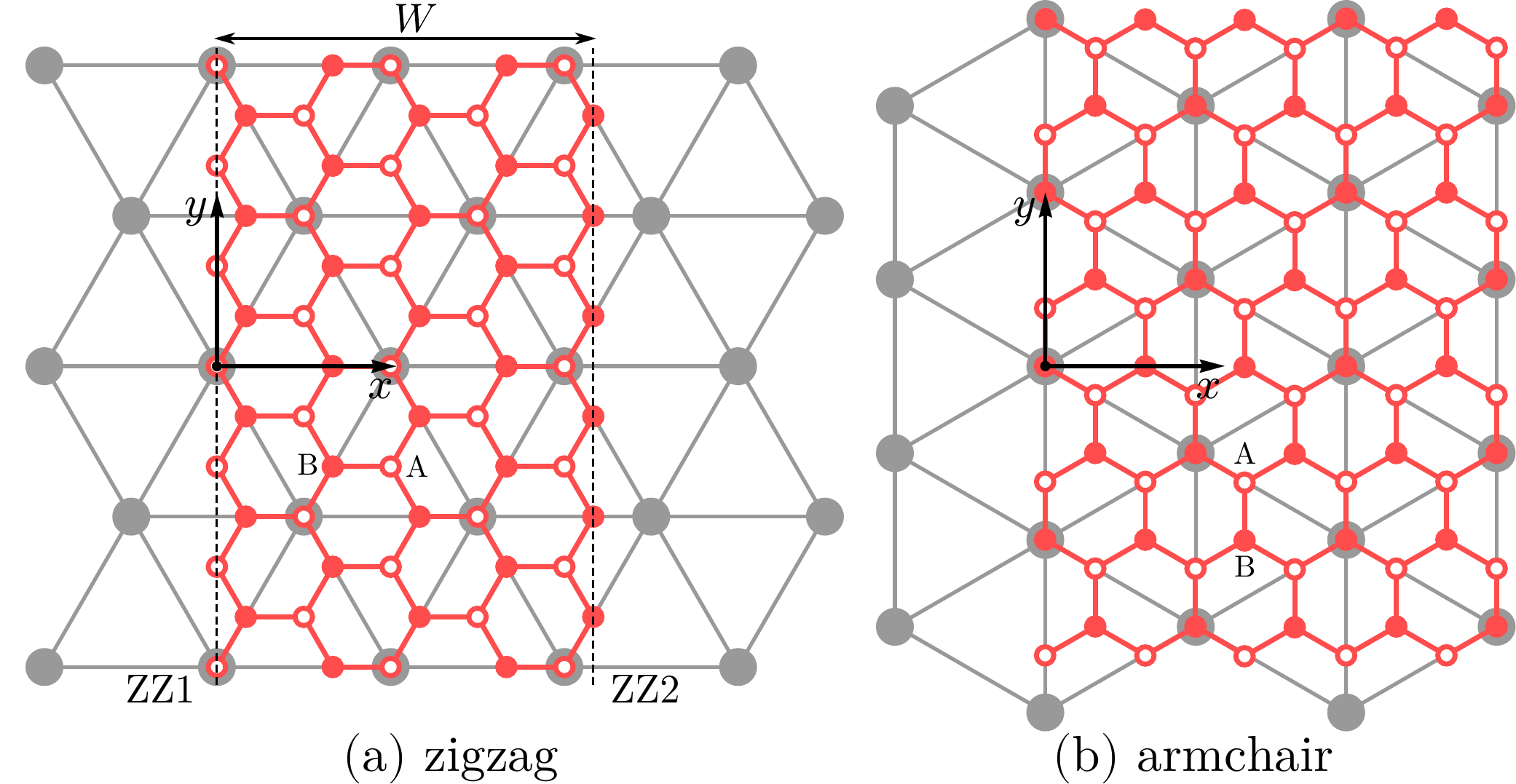}
	\caption{(Color online) Basic edge geometries of graphene (red, small dots) on top of a TIS (gray, large dots) for the T stacking configuration. (a) Zigzag edges: two types depending on whether the edge is terminated by the A (ZZ1) or B (ZZ2) sublattice. (b) One of the three physically distinct armchair edges which the continuum model cannot distinguish.}
	\label{fig:geometry}
\end{figure}

The boundary conditions at $x=0$ are given by the continuity of the TIS spinor components together with the appropriate open boundary conditions for the graphene components depending on the type of edge \cite{bilayer1,bilayer2}. We consider three different edge geometries, shown in Fig.\ \ref{fig:geometry}. For the T structure there are two distinct types of zigzag edges: one terminated by sublattice A (ZZ1) and one terminated by sublattice B (ZZ2). For the armchair edge (AC) there are three different edge configurations, but the continuum model cannot distinguish any of them because the armchair edge contains both sublattices. In case of B stacking, shown in Fig.\ \ref{fig:stacking}, there is also no distinction between the ZZ1 and ZZ2 edges within the continuum model.

The continuity of the TIS spinor components gives
\begin{equation} \label{eq:boundTIS}
\Phi_{I}(0) = \left. \Phi_{II}(0) \right|_{\rm TIS}.
\end{equation}
Next, we consider the boundary conditions for the graphene components. For the zigzag edge, shown in Fig.\ \ref{fig:geometry} (a), the boundary condition is satisfied by putting the spinor component of the relevant sublattice equal to zero at the edge for the two valleys separately \cite{brey2006}. For a zigzag edge at $x=0$, this gives
\begin{equation} \label{eq:boundZZ}
\left. \Phi_{II}(0) \right|_{\alpha{\uparrow(\downarrow)}} = \left. \Phi_{II}(0) \right|_{\alpha'{\uparrow(\downarrow)}} = 0,
\end{equation}
where $\alpha = A, B$ for the ZZ2 and ZZ1 boundary conditions, respectively.
For the armchair edge, shown in Fig.\ \ref{fig:geometry} (b), the boundary condition only yields a nontrivial solution if the $K$ and $K'$ valleys of graphene are coupled by the edge because an armchair edge contains both sublattices \cite{brey2006}.
The boundary condition for the armchair edge is thus given by
\begin{equation}
\left. \Psi_{K} e^{i \bm K \cdot \bm r} + \Psi_{K'} e^{i \bm K' \cdot \bm r} \right|_{\rm edge} = 0,
\end{equation}
where $\Psi_{K}$ and $\Psi_{K'}$ are the graphene spinors.
For the coordinate system shown in Fig.\ \ref{fig:geometry} (b), and $\bm K' = -\bm K = \frac{4\pi}{3a} \bm{\hat x}$ where $a$ is the graphene lattice constant, we have
\begin{equation}
\Psi_{K} = \begin{pmatrix} \psi_A \\ \psi_B \end{pmatrix}, \quad \Psi_{K'} = \begin{pmatrix} \psi_{A'} \\  \psi_{B'} \end{pmatrix},
\end{equation}
for both spin components. Note that we have chosen the Hamiltonian in such a way that no phase factors arise in the components. In the zigzag case, this is of no concern, since relative phase factors between valleys drop out of the boundary condition. Hence, it does not matter that we used rotated coordinates for the zigzag case, as shown in Fig.\ \ref{fig:geometry}(a). Thus, we find that the armchair boundary condition at $x=0$ is given by
\begin{equation}
\label{eq:boundAC}
\left. \Phi_{II}(0) \right|_{\alpha{\uparrow(\downarrow)}} + \left. \Phi_{II}(0) \right|_{\alpha'{\uparrow(\downarrow)}} = 0, \quad \alpha=A,B.
\end{equation}

In general, the combined boundary conditions from Eq.\ \eqref{eq:boundTIS} and Eqs.\ \eqref{eq:boundZZ} or \eqref{eq:boundAC} result in six equations that are solved numerically and yield the reflection coefficient $r$ and the five transmission coefficients $t_n$.

\subsubsection{Transmission channels}

There are five scattering channels in the heterostructure region for the graphene step, while there is only one reflection channel for the bare topological-insulator surface. In order to obtain the transmission probability of the different scattering channels, we consider the probability current in the $x$ direction. The probability-current operator in the $x$ direction is given by
\begin{equation}
j = \left( v_g s_0 \otimes \sigma_x \right) \oplus \left( -v_g s_0 \otimes \sigma_x \right) \oplus \left( -v_s s_y \right).
\end{equation}
By definition, the transmission probability of the $n$th scattering channel is given by
\begin{equation}
T_n = \frac{\psi_n^\dag j \psi_n}{\phi_i^\dag j \phi_i} \left| t_n \right|^2 = \frac{\psi_n^\dag j \psi_n}{2Ek_x} \left| t_n \right|^2,
\end{equation}
and the total transmission probability $T = \sum_{n=1}^5 T_n$. For scattering modes of the valley odd graphene Dirac cones ($E^2>v_g^2k_y^2$), that are decoupled in the bulk, we further obtain from Eq.\ \eqref{eq:oddmodes},
\begin{equation}
T_4 = \frac{2q_{4x}}{k_x} \left| t_4 \right|^2, \qquad T_5 = \frac{2q_{5x}}{k_x} \left| t_5 \right|^2,
\end{equation}
while $T_4 = T_5 = 0$ for evanescent modes ($E^2<v_g^2k_y^2$).
The reflection probability $R$ is given by
\begin{equation}
R = -\frac{\phi_r^\dag j \phi_r}{\phi_i^\dag j \phi_i} \left| r \right|^2 = \left| r \right|^2,
\end{equation}
where conservation of the probability current requires that $R + T = 1$. Before we discuss our results for the step geometry, we consider the boundary conditions for the nanoribbon barrier.

\subsection{Graphene nanoribbon barrier}

Here, we consider a barrier composed of a graphene nanoribbon deposited on the TIS in the T stacking configuration. The ribbon is infinite along the $y$ direction and finite in the $x$ direction with width $W$. This is illustrated for the zigzag barrier in Fig.\ \ref{fig:geometry} (a). 

\subsubsection{Scattering states}

The scattering state of the TIS for $x<0$ is again given by Eq.\ \eqref{eq:sol1}. In the barrier region ($0<x<W$), the wave function can be written as
\begin{equation} \label{eq:sol2barrier}
\Phi_{II}(x) = \sum_{n=1}^5 a_n \psi_{n+} e^{iq_{nx} x} + b_n \psi_{n-} e^{-iq_{nx} x},
\end{equation}
where the wave vectors $q_{nx}$ are found from Eqs.\ \eqref{eq:mom2} and \eqref{eq:mom3} and the spinor $\psi_{n\pm}$ corresponds to $\pm q_{nx}$. Note that we do not need to worry about the correct sign of the wave vector because both are admissible in the finite barrier. Behind the barrier $(x>W)$, the solution becomes
\begin{equation}
\Phi_{III}(x) = t \phi_t e^{ik_x x},
\end{equation}
where $t$ is the reflection coefficient, the spinor $\phi_t=\phi_i$ is given in Eq.\ \eqref{eq:spinor1}, and $k_x$ is given in Eq.\ \eqref{eq:mom1}.

\subsubsection{Boundary conditions}

The boundary conditions of the barrier consist of the continuity of the TIS spinor components and the appropriate open boundary conditions for the graphene spinor components at $x=0$ and $x=W$. The former become
\begin{align} \label{eq:barrierTIS}
\Phi_{I}(0) & = \left. \Phi_{II}(0) \right|_{\rm TIS} \\
\Phi_{III}(W) & = \left. \Phi_{II}(W) \right|_{\rm TIS}.
\end{align}
First, we consider the zigzag ribbon. We take the ZZ1 edge at $x=0$ so that the edge at $x=W$ is automatically ZZ2. In this case, the boundary conditions become
\begin{align} \label{eq:barrierZZ}
\left. \Phi_{II}(0) \right|_{B{\uparrow(\downarrow)}} & = \left. \Phi_{II}(0) \right|_{B'{\uparrow(\downarrow)}} = 0 \\
\left. \Phi_{II}(W) \right|_{A{\uparrow(\downarrow)}} & = \left. \Phi_{II}(W) \right|_{A'{\uparrow(\downarrow)}} = 0.
\end{align}
Analogous to the discussion on the armchair edge above, we find that the boundary conditions for the armchair ribbon are given by
\begin{align} \label{eq:barrierAC}
\left. \Phi_{II}(0) \right|_{\alpha{\uparrow(\downarrow)}} + \left. \Phi_{II}(0) \right|_{\alpha'{\uparrow(\downarrow)}} & = 0 \\
\left. \Phi_{II}(W) \right|_{\alpha{\uparrow(\downarrow)}} + e^{i \Delta K \, W} \left. \Phi_{II}(W) \right|_{\alpha'{\uparrow(\downarrow)}} & = 0,
\end{align}
for $\alpha=A,B$, where $\Delta K = 8\pi/3a$.

The boundary conditions for the barrier give twelve equations that are solved numerically and yield the reflection coefficient $r$, the ten barrier coefficients $a_n$ and $b_n$, and the transmission coefficient $t$.

\subsubsection{Bound states}

States of the TIS for which $E^2 < v_s^2 k_y^2$ are evanescent and as such we can have bound states, localized in the graphene nanoribbon. In this case, the wave functions outside the ribbon become
\begin{equation} \label{eq:bound}
\Phi_I(x) = c \begin{pmatrix} E/v_s \\ k_y - \kappa \end{pmatrix} e^{\kappa x}, \quad \Phi_{III}(x) = d \begin{pmatrix} E/v_s \\ k_y + \kappa \end{pmatrix} e^{-\kappa x},
\end{equation}
where $\kappa = \sqrt{k_y^2-(E/v_s)^2}$ and the wave function inside the ribbon is given by Eq.\ \eqref{eq:sol2barrier}. The boundary conditions and the normalization give twelve independent equations for the coefficients $a_n$, $b_n$, $c$, and $d$.

\section{Results} \label{section:results}

In this section, we discuss our numerical results for transmission through a graphene step and nanoribbon barrier deposited on the TIS in the T stacking configuration. We always put $v_s = v_g/2$, which is representative for the TIs listed in Table \ref{tab:materials}, and we present our results for $t=0.3$~eV as an example, unless stated explicitly.
\begin{figure*}
	\centering
	\includegraphics[width=0.8\linewidth]{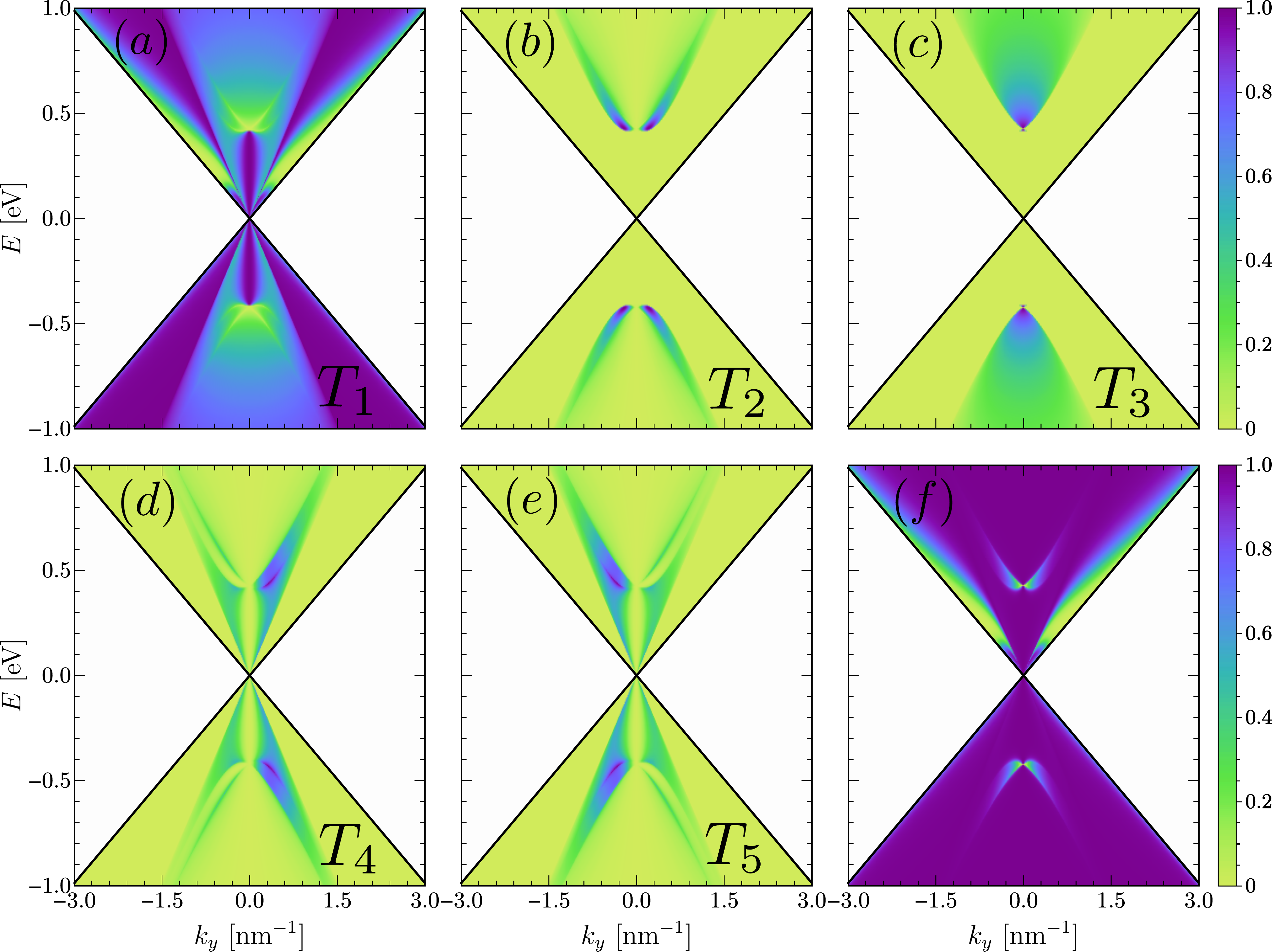}
	\caption{(Color online) (a)-(e) Transmission probabilities $T_n$ for scattering at the ZZ1 step with $t=0.3$~eV for the scattering channels $n=1,\ldots,5$, respectively, and (f) the total transmission probability $T = \sum_{n=1}^5 T_n$.}
	\label{fig:step}
\end{figure*}

\subsection{Graphene step}

Out of the three edges we have considered for the graphene step, only one of the zigzag edges, ZZ1, shows interesting features in the transmission probability $T(E,k_y)$. Interestingly, the result for the ZZ2 and AC edges is exactly the same and shows near perfect transmission, even at oblique angles.
As seen in Fig.\ \ref{fig:geometry}, only for the ZZ1 boundary does the terminated graphene edge couple directly to the TIS lattice. We find that only the ZZ1 edge induces coupling to the valley odd cones that are decoupled for the bulk heterostructure. The transmission probability of the different scattering channels at the ZZ1 edge is shown in Fig.\ \ref{fig:step}, together with the total transmission probability. For $E \lesssim \sqrt{2} t$, the main transmission channel is $T _1$, and the ZZ1 edge allows for some transmission to channels $4$ and $5$, corresponding to the valley odd cones.
At higher energies, the Rashba channels $T_2$ and $T_3$ become available and  the transmission via $T_1$ reduces inside the region $E^2<v_g^2k_y^2$ defined by the graphene Dirac cone. Interestingly, the channels $T_4$ and $T_5$, which are $s_z$ eigenstates and completely localized in graphene for the bulk heterostructure, are mirrored with respect to each other about $k_y=0$. Moreover, they show a preference for either left or right moving states for both electrons and holes, creating a bulk spin-momentum locked state in the graphene originating from valley odd states that are decoupled in the bulk. Note that only $T_1$, and therefore also the total transmission probability, is not symmetric with respect to zero energy. This asymmetry originates from the fact that a step graphene-TIS system has only one interface which breaks the symmetry of the lattice structure, resulting in an asymmetric transmission for electrons and holes, in contrast to the graphene-TIS barrier structure.

\subsection{Graphene nanoribbon barrier}

Now we discuss our results for the transmission across the graphene nanoribbon. The results for the barrier are symmetric with respect to zero energy and we only show the results for positive energy. The width of the graphene ribbons, including dangling bonds, is given by
\begin{align}
W_{ZZ} & = \frac{a}{2\sqrt{3}} \left( 3N + 2 \right), \\
W_{AC} & = \frac{a}{2} \left( N + 1 \right),
\end{align}
where $a$ is the graphene lattice constant and $N$ is the number of two-atom unit cells along the finite $x$ direction.

\begin{figure}
	\centering
	\includegraphics[width=\linewidth]{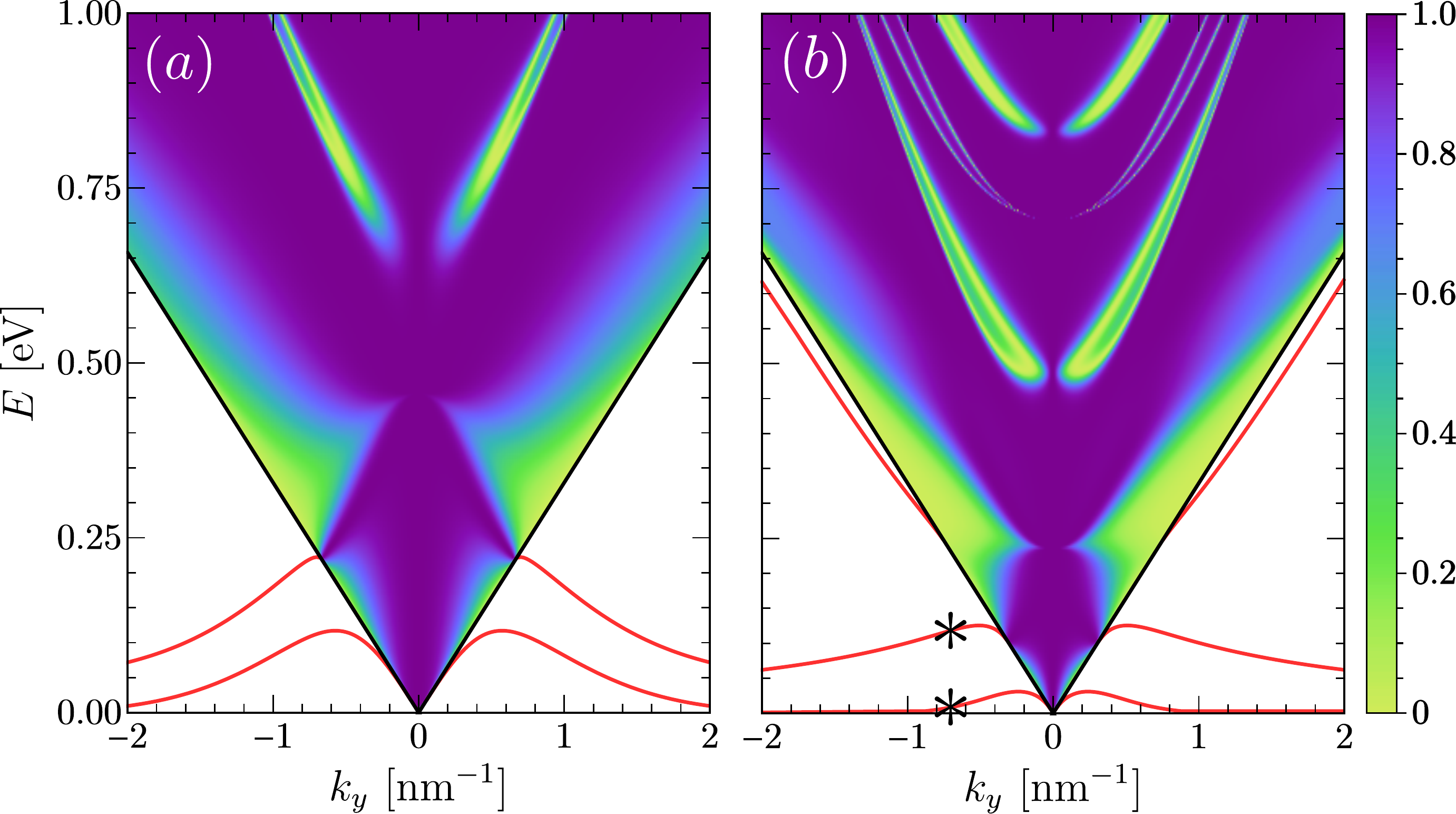}
	\caption{(Color online) Transmission probability $T(E,k_y)$ for a zigzag ribbon with $t=0.3$ eV, and (a) $N=10$ and (b) $N=20$. The red lines outside the cone are bound states and the density corresponding to the states marked with an asterix is shown in Fig.\ \ref{fig:wave}.}
	\label{fig:barrier1}
\end{figure}

\begin{figure}
	\centering
	\includegraphics[width=\linewidth]{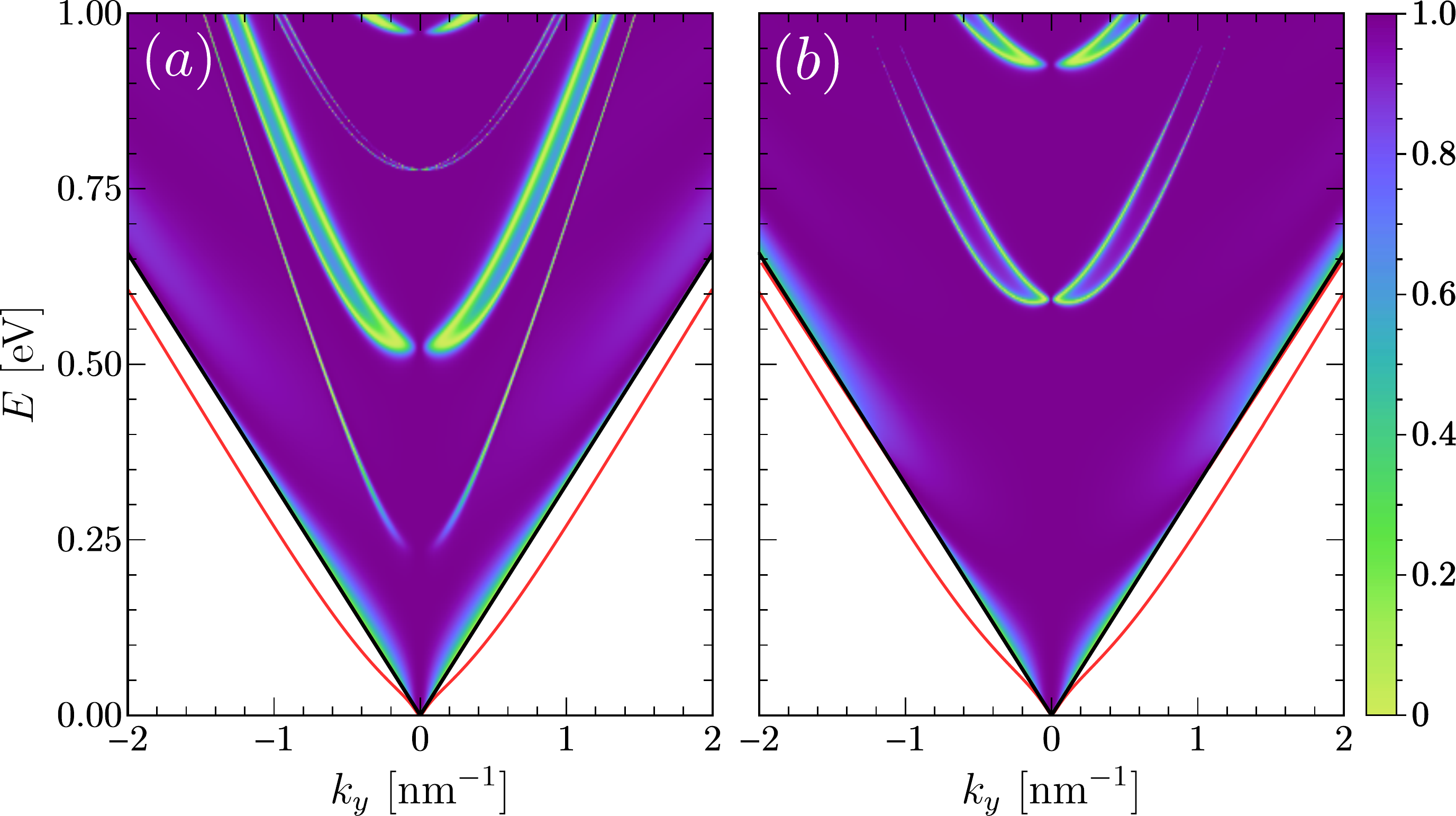}
	\caption{(Color online) Transmission probability $T(E,k_y)$ for an armchair ribbon with $t=0.3$ eV, and (a) $N=30$ (insulating) and (b) $N=41$ (metallic). The red lines outside the cone are bound states.}
	\label{fig:barrier2}
\end{figure}

In Figs.\ \ref{fig:barrier1} and \ref{fig:barrier2}, we show the transmission probability for the zigzag and armchair barrier, respectively. The transmission probability is always equal to unity at normal incidence for both zigzag and armchair ribbons, which is what we expect for a nonmagnetic scatterer on the TIS \cite{}. Moreover, we observe two resonances at low energies for the zigzag ribbon and antiresonances for both the zigzag and armchair ribbons. The low-energy resonances for the zigzag ribbons, shown in Fig.\ \ref{fig:barrier1}, are caused by edge states, that are absent for an armchair ribbon.
\begin{figure}
	\centering
	\includegraphics[width=\linewidth]{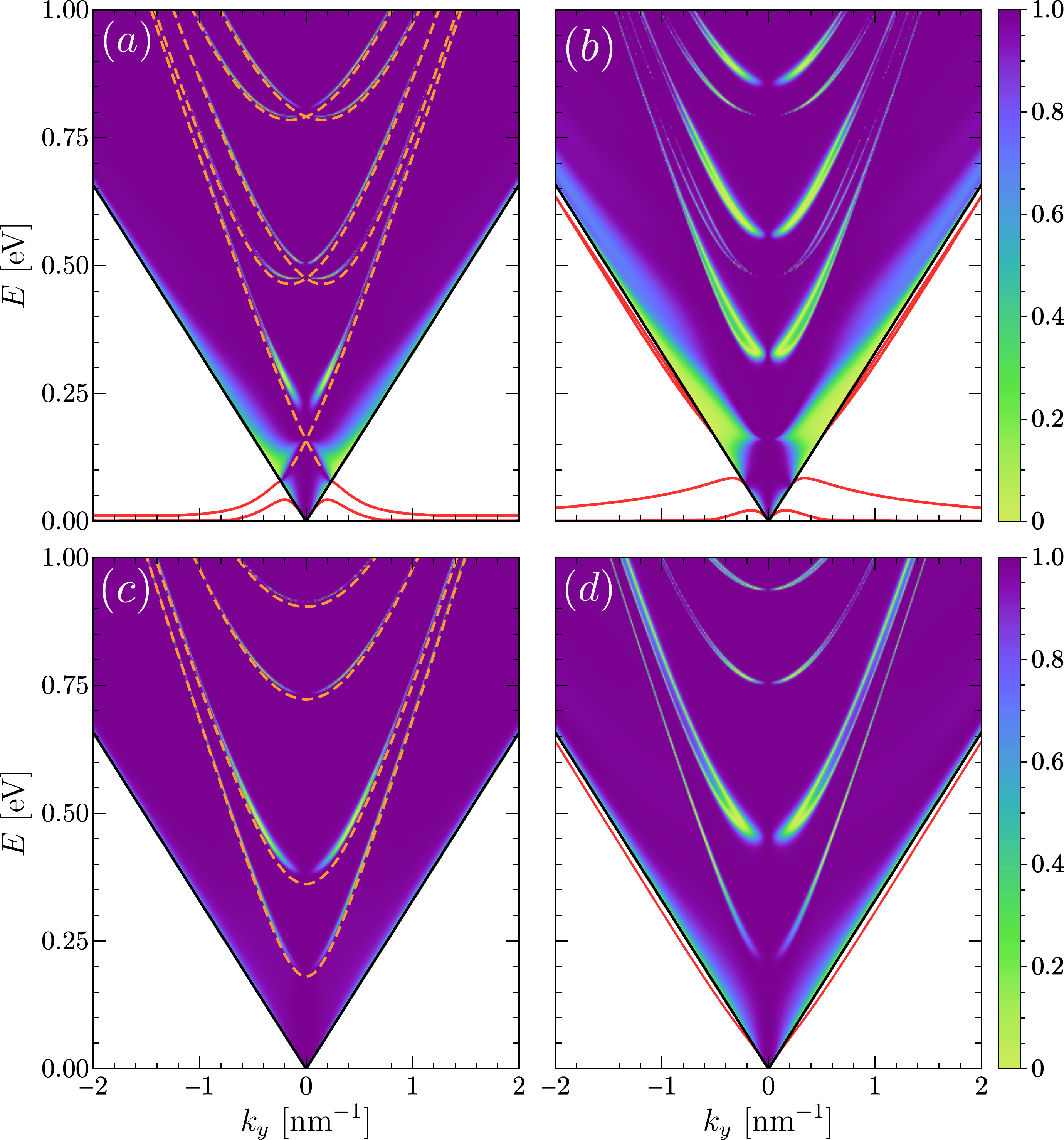}
	\caption{(Color online) Transmission probability $T(E,k_y)$ for the nanoribbon barrier. (a)-(b) Zigzag ribbon with $N=30$ for (a) $t=0.1$ eV and (b) $t=0.2$ eV. (c)-(d) Armchair ribbon with $N=30$ for (c) $t=0.1$ eV and (d) $t=0.2$ eV. The red lines are bound states, localized in the barrier, while the orange dashed lines in (a) and (c) are the bound states of a bare graphene nanoribbon.}
	\label{fig:barrier3}
\end{figure}

\begin{figure}
	\centering
	\includegraphics[width=\linewidth]{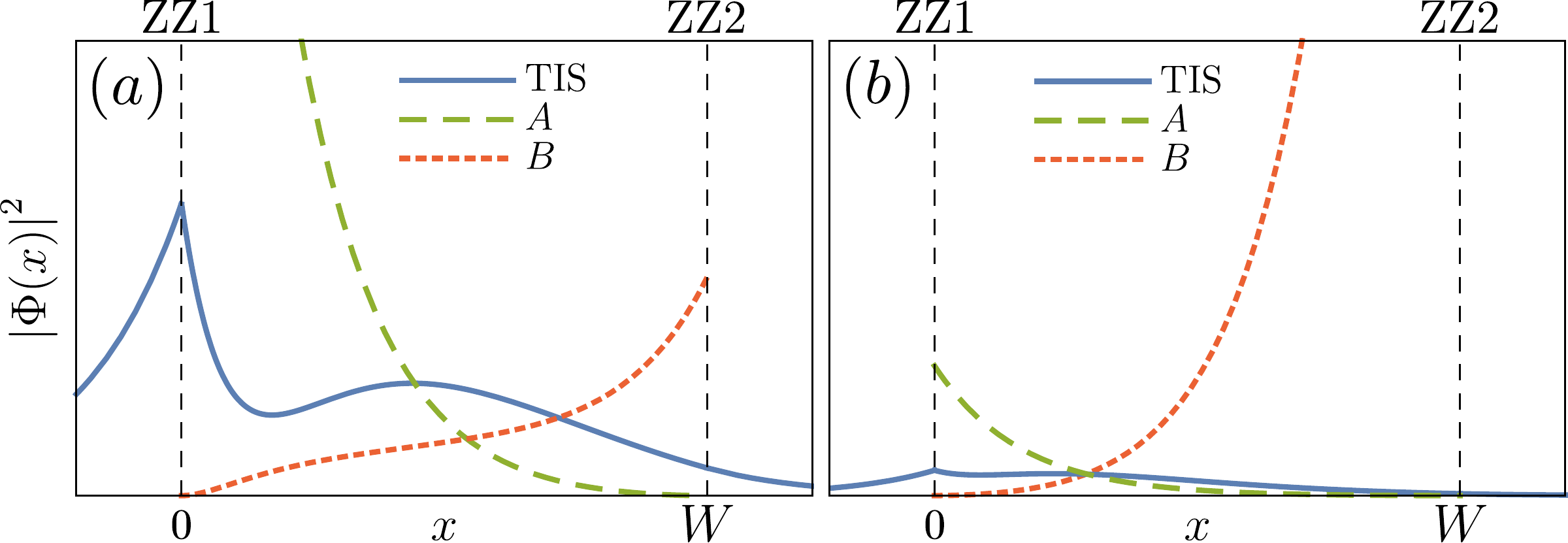}
	\caption{(Color online) Projected electron density of the (a) lower and (b) upper branch of edge states for the zigzag ribbon with $N=20$ and $t=0.3$ for $k_y = 0.7$ nm$^{-1}$. These states are marked in Fig.\ \ref{fig:barrier1} with an asterisk.}
	\label{fig:wave}
\end{figure}

\begin{figure}
	\centering
	\includegraphics[width=0.7\linewidth]{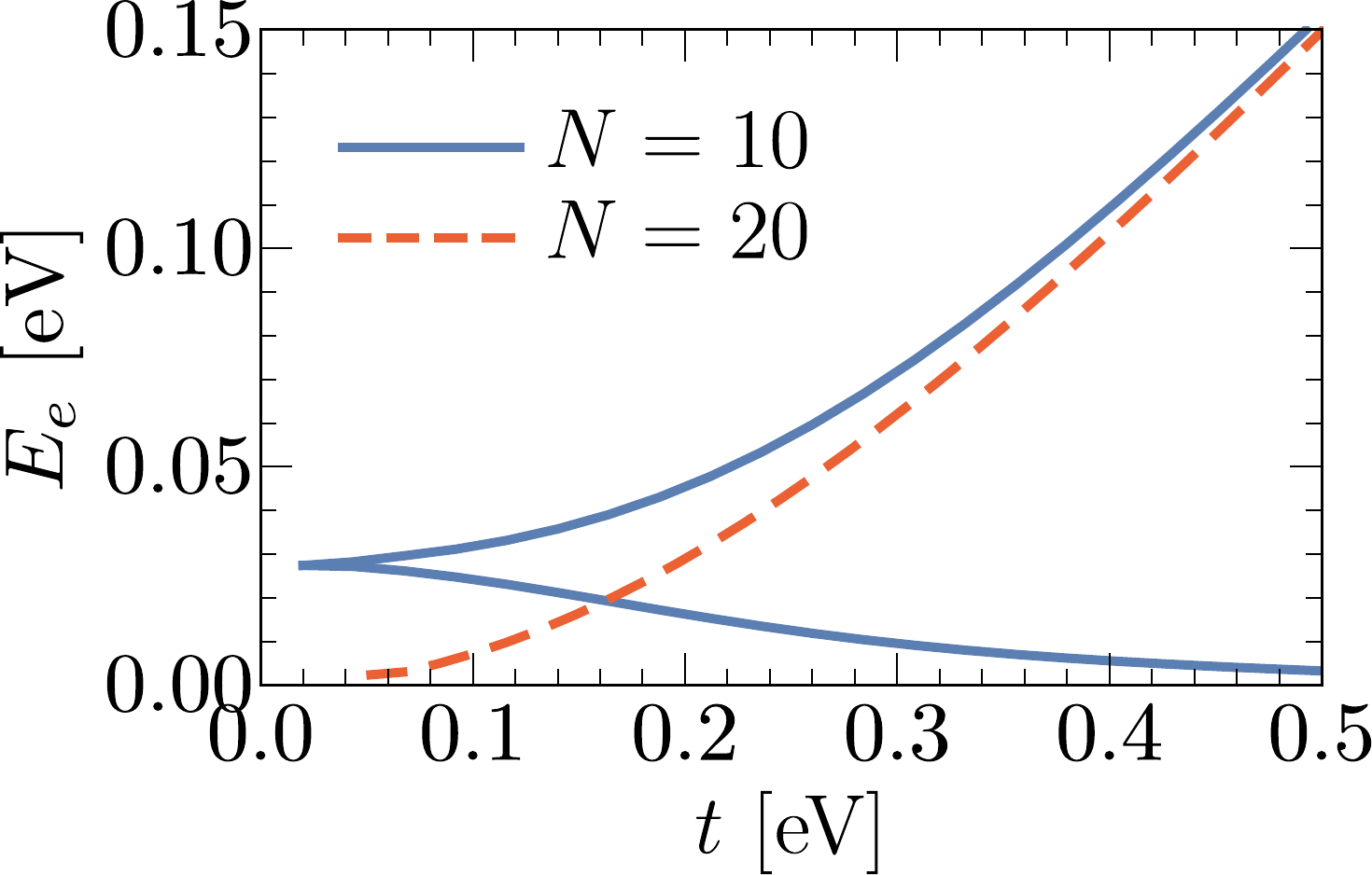}
	\caption{(Color online) Energy of the zigzag edge states at $k_y=2$ nm$^{-1}$ as a function of $t$ for $N=10$ and $N=20$. We only show one state for $N=20$, since the other state remains at zero energy for all $t$.}
	\label{fig:splitting}
\end{figure}

To understand the nature of these edge states and the antiresonances, we consider the evolution of the transmission probability as a function of the coupling $t$ between graphene and the TIS. In Fig.\ \ref{fig:barrier3}, we plot the transmission probability for (a)-(b) zigzag and (c)-(d) armchair ribbons with $t=0.1$~eV and $t=0.2$~eV. We see that the two positive-energy edge states for the zigzag ribbon split with increasing $t$. The upper branch is localized on the ZZ1 edge which couples directly to the TIS, while the lower branch is localized on the ZZ2 edge which has no direct coupling to the TIS. In Fig.\ \ref{fig:wave}, we show the electron density for a fixed value of $k_y$ for both edge states corresponding to Fig.\ \ref{fig:barrier1} (b). Note that the upper branch is actually a hybridized state of graphene and the TIS, localized near the ZZ1 edge.
The energy splitting of the edge states is shown in Fig.\ \ref{fig:splitting} as a function of $t$ for $N=10$ and $N=20$. For $N=10$, there is a confinement effect near $t=0$ which is absent for $N=20$. However, this confinement splitting is lifted when $t$ increases because the energy difference of states localized at different edges increases, and the lower branch returns to zero energy. The energy of the upper branch grows linearly with $t$, since the coupling with the TIS splits the two formerly $s_z$ eigenstates localized on the ZZ1 edge. Moreover, if the barrier is wide enough or the coupling strong enough, there are also bound states that are delocalized over the entire ribbon, both in the zigzag and armchair case, as is shown in Figs.\ \ref{fig:barrier1}, \ref{fig:barrier2}, and \ref{fig:barrier3}.

Furthermore, in Fig.\ \ref{fig:barrier3} (a) and (c), we have superimposed the bound states of a bare graphene ribbon on the transmission probability for $t=0.1$ eV for both an armchair and zigzag barrier. In this case, the antiresonances are very sharp and coincide almost perfectly with the bound states of the bare ribbon. These antiresonances are quasibound states originating from both valley even and valley odd states. With increasing $t$, the quasibound states split into two classes: Those that broaden and move in energy with increasing $t$ correspond to the Rashba-split bands while those that remain very sharp and almost at the same energy correspond to the valley odd cones. Indeed, the latter are missing for the armchair barrier because the AC edge does not induce coupling to these states. Note that the coupling due to the ZZ1 edge also induces some spin splitting into the quasibound states originating from the valley odd states. At these energies, the wave function is either strongly hybridized, which is the case for the Rashba-like states, or completely localized in the graphene, which is the case for the valley odd states. In the latter case, which only occurs for zigzag ribbons, tunneling is impossible since the ribbon contains at least one edge that does not allow tunneling to these states. On the other hand, the Rashba-like bound states of the graphene ribbon, induced by the ribbon confinement, can only lead to more possibilities for backscattering, and thus antiresonances.
In Figs.\ \ref{fig:barrier1} and \ref{fig:barrier2} the antiresonances are broadened compared to Fig.\ \ref{fig:barrier3} because the coupling  to the TIS is stronger.

\subsubsection*{Conductance}

\begin{figure}
	\centering
	\includegraphics[width=\linewidth]{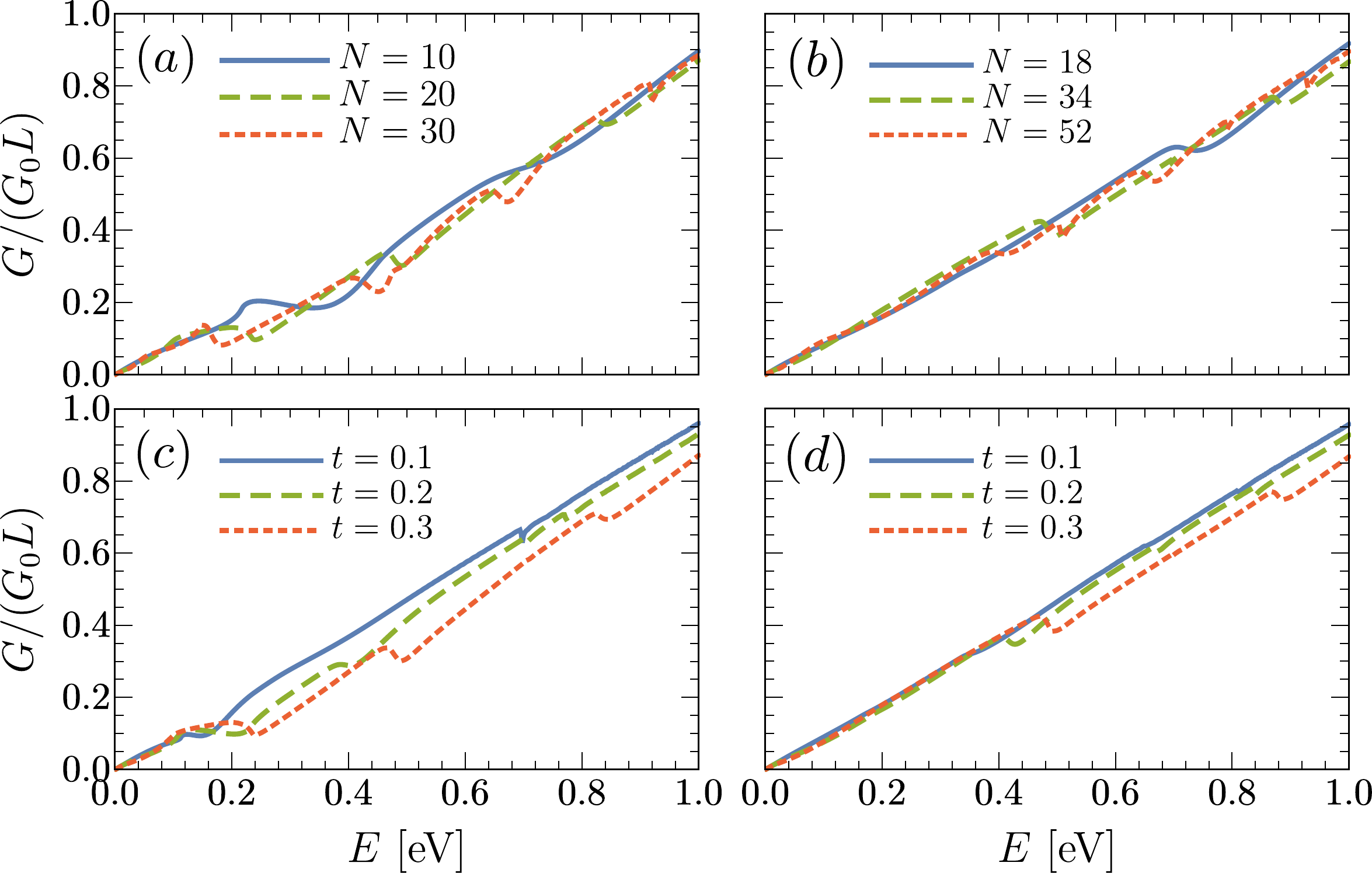}
	\caption{(Color online) (a)-(b) Conductance for the (a) zigzag and (b) armchair barrier for several widths with $t=0.3$~eV. The widths of the armchair ribbon are chosen so that it is insulating and matches the corresponding widths in the zigzag case. (c)-(d) Conductance for the (c) zigzag barrier with $N=20$ and (d) armchair barrier with $N=34$ for several $t$, whose values are shown in eV.}
	\label{fig:cond}
\end{figure}
The zero-temperature conductance through a barrier of width $W$ and length $L$ is given by
\begin{equation}
G(E) = G_0 \, \frac{L}{2\pi} \int_{-\frac{|E|}{\hbar v_s}}^{\frac{|E|}{\hbar v_s}} dk_y \, T(E,k_y),
\end{equation}
where $G_0 = 2e^2/h$ is the conductance quantum and where we have used dimensionful units. This is a weighted sum over the available incident transverse modes $L|E|/\left(\pi \hbar v_s\right)$. The conductance for zigzag and armchair graphene nanoribbons deposited on the TIS in the T stacking configuration are shown in Fig.\ \ref{fig:cond} for several values of the width $W$ and the coupling $t$.

The plateaus in the conductance are caused by the antiresonances in the transmission probability discussed above. They are more pronounced for the zigzag barrier than the armchair barrier. With increasing $N$, the number of plateaus increase and they move towards zero energy because of the reduced confinement. On the other hand, if we increase $t$, more plateaus appear in the conductance and it is suppressed overall due to backscattering at oblique angles.

\section{Summary and conclusions} \label{section:summary}

In summary, we have considered the electronic transmission, using a continuum model, of the topological surface state of a three dimensional time-reversal invariant topological insulator
through heterostructures made by depositing a monolayer graphene on the topological-insulator surface. We obtained the transmission of the topological surface state through a semi-infinite graphene step and a graphene nanoribbon for both zigzag and armchair boundaries. We found that the transmission depends strongly on the type of edge: In the case of a graphene step, we found that the transmission exhibits electron-hole asymmetry for the  ZZ1 edge configuration while the transmission is perfect at all energies for armchair and ZZ2 junctions. Moreover, our results show that the conductance through a graphene nanoribbon exhibits plateaus caused by antiresonances in the transmission probability at energies of the quasibound states of the deposited nanoribbon for both zigzag and armchair edges. 

The heterostructures we considered are commensurate by less than one percent with at least two well-known topological insulators, Sb$_2$Te$_3$ and TlBiSe$_2$. Hybrid graphene-TI devices could be fabricated using a mechanical transfer method where the chemical potential difference and electron density can be tuned by gate voltages. Further studies are required to address the effect of an external magnetic field and the number of graphene layers on the transport properties.

\begin{acknowledgments}
This research was supported by the Flemish Research Foundation (FWO).
We thank B.\ Van Duppen for interesting discussions.
\end{acknowledgments}

\appendix

\section{Unitary transformation}

Here, we give the explicit expression for the unitary transformation $U_{\bm k}$ that block diagonalizes the Hamiltonian from Eq.\ \eqref{eq:ham} for the case $t_B = 0$ (T structure), into the form shown in Eqs.\ \eqref{eq:block}, \eqref{eq:block1}, and \eqref{eq:block2}. We find
\begin{equation}
U_{\bm k} =
\begin{pmatrix}
A_{\bm k} & B_{\bm k} & 0 \\
A_{\bm k} & -B_{\bm k} & 0 \\
0 & 0 & 1
\end{pmatrix},
\end{equation}
with
\begin{align}
A_{\bm k} & = \frac{1}{\sqrt{2}} \diag \left( 1, -e^{-2i\theta_{\bm k}}, 1, 1 \right) \\
B_{\bm k} & = \frac{1}{\sqrt{2}} \diag \left( 1, 1, 1 , -e^{2i\theta_{\bm k}} \right),
\end{align}
where $\theta_{\bm k} = \arctan(k_y/k_x)$.

After performing this unitary transformation, the new basis states are
\begin{align}
\left| \psi_{A^\pm{\uparrow(\downarrow)}} \right> & = \frac{1}{\sqrt{2}} \left( \left| \psi_{A{\uparrow(\downarrow)}} \right> \pm \left| \psi_{A'{\uparrow(\downarrow)}} \right) \right> \\
\left| \psi_{B^\pm{\uparrow}} \right> & = \frac{1}{\sqrt{2}} \left( \mp e^{-2i\theta_{\bm k}} \left| \psi_{B{\uparrow}} \right> + \left| \psi_{B'{\uparrow}} \right> \right) \\
\left| \psi_{B^\pm{\downarrow}} \right> & = \frac{1}{\sqrt{2}} \left( \left| \psi_{B{\downarrow}} \right> \mp e^{2i\theta_{\bm k}} \left| \psi_{B'{\downarrow}} \right> \right),
\end{align}
where $\pm$ corresponds to the even and odd valley subspace. Under time reversal, the new basis transforms as
\begin{equation}
\begin{aligned}
\Theta \left| \psi_{A^\pm{\uparrow(\downarrow)}} \right> & = \pm \left| \psi_{A^\pm{\downarrow(\uparrow)}} \right> \\
\Theta \left| \psi_{B^\pm{\uparrow(\downarrow)}} \right> & = \left| \psi_{B^\mp{\downarrow(\uparrow)}} \right>,
\end{aligned}
\end{equation}
so that
\begin{equation}
\begin{aligned}
\left< \psi_{A^-{\uparrow(\downarrow)}} \right| V \left| \phi_{\uparrow(\downarrow)} \right> & = \left< \psi_{A^-{\uparrow(\downarrow)}} \right| \Theta^{-1} V \Theta \left| \phi_{\uparrow(\downarrow)} \right> \\
& = - \left< \psi_{A^-{\downarrow(\uparrow)}} \right| V \left| \phi_{\downarrow(\uparrow)} \right> \\
& = - \left< \psi_{A^-{\uparrow(\downarrow)}} \right| V \left| \phi_{\uparrow(\downarrow)} \right> = 0,
\end{aligned}
\end{equation}
and the matrix element between the odd subspace and the topological surface state vanish.

\end{document}